\documentclass[aps,prl,tighetenlines,notitlepage,superscriptaddress,showpacs]{revtex4-1}
\usepackage{graphicx}
\usepackage{amssymb}
\usepackage{amsmath}
\usepackage{epstopdf}
\usepackage{natbib}
\usepackage{bm}

\begin{document}
\title{Motion of magnetotactic bacteria swarms in an external field}
\author{K.Bente}
\affiliation{Department of Biomaterials, Max Planck Institute of Colloids and Interfaces, Science Park Golm, 14424 Potsdam, Germany}
\author{G.Kitenbergs}
\affiliation{MMML lab, Department of Physics, University of Latvia, Ze\c{l}\c{l}u 25, R\={\i}ga, LV-1002, Latvia}
\author{D.Krimans}
\affiliation{MMML lab, Department of Physics, University of Latvia, Ze\c{l}\c{l}u 25, R\={\i}ga, LV-1002, Latvia}
\author{K.\={E}rglis}
\affiliation{MMML lab, Department of Physics, University of Latvia, Ze\c{l}\c{l}u 25, R\={\i}ga, LV-1002, Latvia}
\author{M.Belovs}
\affiliation{MMML lab, Department of Physics, University of Latvia, Ze\c{l}\c{l}u 25, R\={\i}ga, LV-1002, Latvia}
\author{D.Faivre}
\affiliation{Department of Biomaterials, Max Planck Institute of Colloids and Interfaces, Science Park Golm, 14424 Potsdam, Germany}
\author{A.C\={e}bers} \email[]{aceb@tesla.sal.lv}
\affiliation{MMML lab, Department of Physics, University of Latvia, Ze\c{l}\c{l}u 25, R\={\i}ga, LV-1002, Latvia}

\date{\today}
\begin{abstract}
Magnetotactic bacteria moving on circular orbits form hydrodynamically bound states. When close to a surface and with the tilting of the field in a direction close to the perpendicular to this surface these swarms move perpendicularly to the tilting angle. We describe quantitatively this motion by a continuum model with couple stress arising from the torques produced by the rotary motors of the amphitrichous magnetotactic bacteria. The model not only correctly describes the change of direction of swarm motion while inverting the tangential field but also predicts reasonable value of the torque produced by the rotary motors. 
\end{abstract}


\maketitle

To achieve the full scientific and technological potential of microswimmers, current major challenges include understanding their behavior in complex and crowded environments and learning to engineer emergent behaviors while providing a propulsion mechanism with directional control \cite{1}. This  may be achieved efficiently by torques acting on magnetic microswimmers, such as magnetotactic bacteria, in an applied magnetic field.

Torques applied by magnetic fields cause different hydrodynamic phenomena in ensembles of magnetic particles: an increase of the effective viscosity of magnetic liquids under the action of the external field \cite{2}, macroscopic motion of ferrofluid with free boundaries in a rotating field \cite{3} and many others. These phenomena are described by models of so-called hydrodynamics with spin, which considers the antisymmetric stress determined by the acting volume density of torques. These models also allow the  description of different phenomena in bacterial suspensions such as rotating crystals \cite{4} or  the large self-propulsion velocities of some bacterial species \cite{5} to name a few.

Antisymmetric stress may appear in bacterial suspensions where each bacterium creates a torque dipole on the liquid. One torque is applied by a rotating flagellum  driven by the rotary motor and the second in the opposite direction is due to the rotation of the body of the bacterium. In the case of a suspension with an orientational order, the torque dipoles cause a couple stress and a flow of the suspension.

\begin{figure}[h!]
\includegraphics[width=0.65\textwidth]{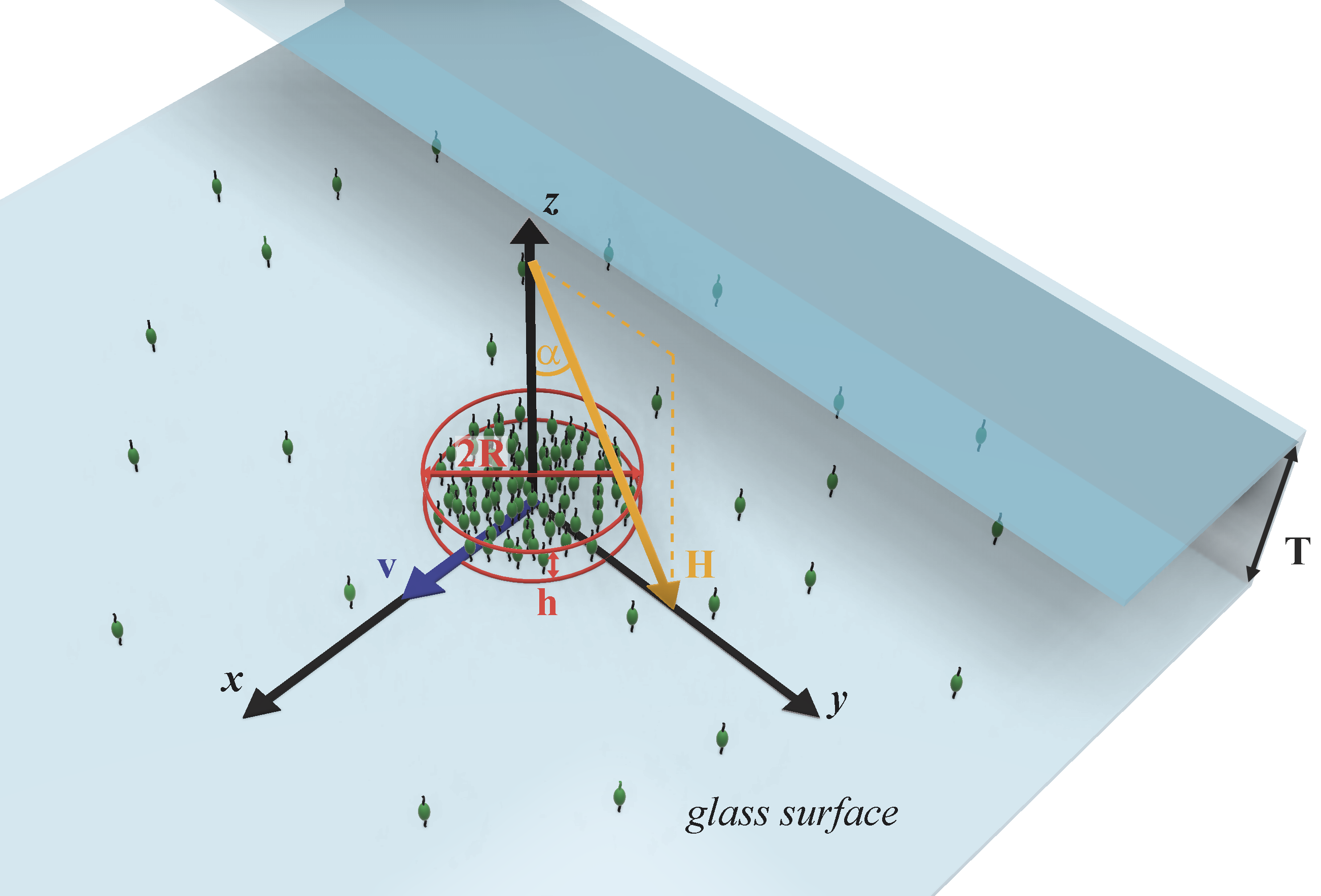}
\caption{Schematic of the investigated system. A swarm of magnetic bacteria with a diameter $2R$ and height $h$ is formed on the bottom glass surface of a capillary with thickness $T$ due to magnetic field $H$ component that is perpendicular to the surface. A small deviation of the field by angle $\alpha$ induces a motion with velocity of swarm $v$ in the direction orthogonal to the plane of the field.}
\label{Fig:00}
\end{figure}

This phenomenon arises in a suspension of magnetotactic bacteria (MTB) in an external field. We utilized the magnetotactic bacteria \textit{Magnetospirillium gryphiswaldense} (MSR-1). These bacteria feature magnetic nanoparticles which are ordered in a chain along the long axis of these helically shaped organisms. Thus, each cell features a magnetic dipole \cite{6,7}. This holds the biological advantage of a simplification of their navigation task towards certain oxygen concentrations along  magnetic field of the Earth \cite{8,9}. We created swarms of MTB with orientation order by concentrating them in a micro-capillary and applying a magnetic field perpendicular to the capillary wall. When the field was inclined, these swarms start to move perpendicularly to their propulsion direction (the inclination direction of the field), as indicated in Figure \ref{Fig:00}. We remark here that this character of motion differs from the motion of bacterial swarms in the direction of the force applied by  bacteria stuck  in a confined geometry \cite{10}.  

We presume that the unusual motion of swarms perpendicular to their ordered propulsion direction is caused by the couple stress in this layer of bacteria near the capillary walls. Couple stress in an oriented ensemble of bacteria may be calculated as in \cite{11}. We approximate the torques acting on the liquid by two torques $\vec{\tau}=\mp\tau\vec{l}$ along the axis of a bacterium $\vec{l}$ and applied at points at distance $b$. Let the volume density of bacteria be $n$ and consider the torque acting on the surface element with the normal $\vec{n}$ and area $\Delta S$ of the material volume of the suspension. It is clear that the torque on the material volume is created only by the bacteria whose axes intersect the surface element. Then for the torque created by all bacteria, whose axes with orientation along $\vec{l}$ intersects the surface element, we have $\tau\vec{l}nb(\vec{l}\cdot\vec{n})\Delta S$ , where $nb(\vec{l}\cdot\vec{n})\Delta S$ is the total number of  such bacteria. The surface density of this torque gives the couple stress $\vec{j}_{k}=\tau nb\vec{l}l_{k}$.

From the equation of motion and the conservation of angular momentum (we neglect the angular momentum of the bacteria due to their small size) then follows
\begin{equation}
\sigma^{a}_{i}=e_{ilm}\sigma_{lm}=\frac{\partial j_{ik}}{\partial x_{k}}~.
\label{Eq:1}
\end{equation}
Thus, a non-uniform distribution of couple stress induces stress in the liquid and, as a result, the liquid moves.

We are considering the flow arising due to the couple stress of uniformly distributed bacteria in a cylindrical region with a radius $R$ and a height $h$. This region lies on the solid surface where a no-slip condition for the velocity of the suspension is applied. The viscosities of the bacterial suspension and the surrounding liquid are assumed to be equal for simplicity.

The relation (\ref{Eq:1}) for the given couple stress distribution $\vec{j}_{k}=\tau nb\vec{l}l_{k}\theta(R-r)\theta(h-z)$ gives  ($\theta$ is the Heaviside function, $z$ axis is perpendicular to the layer, $\vec{l}=(0,l_{y},l_{z})$)
\begin{equation}
\vec{\sigma}^{a}=-\tau nb \vec{l}(\vec{l}\cdot\vec{e}_{r})\delta(R-r)\theta(h-z)-\tau nb\vec{l}l_{z}\theta(R-r)\delta(h-z)~.
\label{Eq:2}
\end{equation}

Since we are interested in the symmetry breaking flow, we will further consider only terms in relation (\ref{Eq:2}) proportional to $l_{y}l_{z}$.

The three dimensional flow in the semi-space above the solid wall at $z=0$ is described by the equation of motion

\begin{equation}
-\nabla p+\eta\Delta\vec{v}+\frac{1}{2}\nabla\times\vec{\sigma}^{a}=0~.
\label{Eq:3}
\end{equation}

The solution of Eq.(\ref{Eq:3}) is found by utilizing the Fourier transform in the $x,y$ plane according to ($\vec{\rho}=(x,y,0)$)

\begin{equation}
\vec{v}=\frac{1}{2\pi}\int\vec{v}_{\vec{k}}\exp{(i\vec{k}\cdot\vec{\rho})}d\vec{k};~\vec{v}_{\vec{k}}=\frac{1}{2\pi}\int\vec{v}\exp{(-i\vec{k}\cdot\vec{\rho})}d\vec{\rho}
\label{Eq:4}
\end{equation}

and similarly for the pressure.

Long but a straightforward calculation for the $x$-component of the velocity at $z<h$ gives ($\tilde{h}=h/R;~\tilde{x}=x/R;~\tilde{y}=y/R;~\tilde{z}=z/R;~r=\sqrt{\tilde{x}^{2}+\tilde{y}^{2}}$, tildes further are omitted, $J_{0,1,2}$ are the Bessel functions)

\begin{eqnarray}
v_{x}=\frac{nb\tau}{4\eta}l_{y}l_{z}\Bigl(\int^{\infty}_{0}J_{1}(u)J_{0}(ur)\Bigl((1-\exp{(-zu)}
-3\exp{(-hu)}\sinh{(zu)}~~~~~~\\ \nonumber
-zu\exp{(-(h+z)u)}\Bigr)du\\ \nonumber
+\frac{x^{2}-y^{2}}{r^{2}}\int^{\infty}_{0}J_{1}(u)J_{2}(ur)\Bigl(zu\exp{(-(h+z)u)}+1-\exp{(-zu)}-\exp{(-hu)}\sinh{(zu)}\Bigr)du\Bigr)~.
\end{eqnarray}
Introducing the characteristic velocity $v_{0}$ the last relation reads ($\alpha$ is the angle between  the applied field and the normal to the wall)
\begin{equation}
v_{0}=\frac{nb\tau}{4\eta}l_{y}l_{z}=\frac{nb\tau}{8\eta}\sin{(2\alpha)};~v_{x}=v_{0}f(x,y,z,h)~.
\label{Eq:10}
\end{equation}

\begin{figure}
\centering{
\includegraphics[height=5cm]{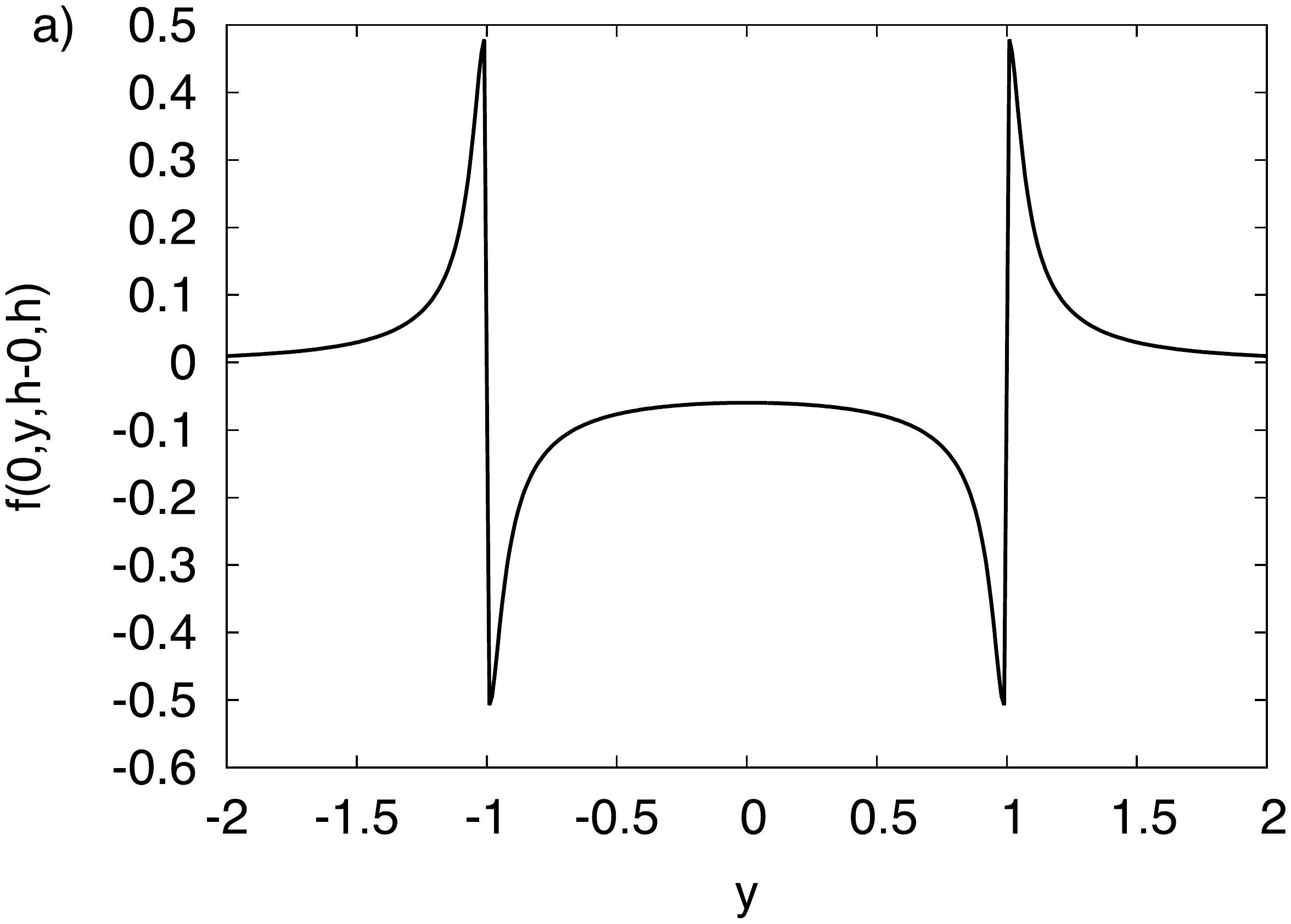}
\includegraphics[height=5cm]{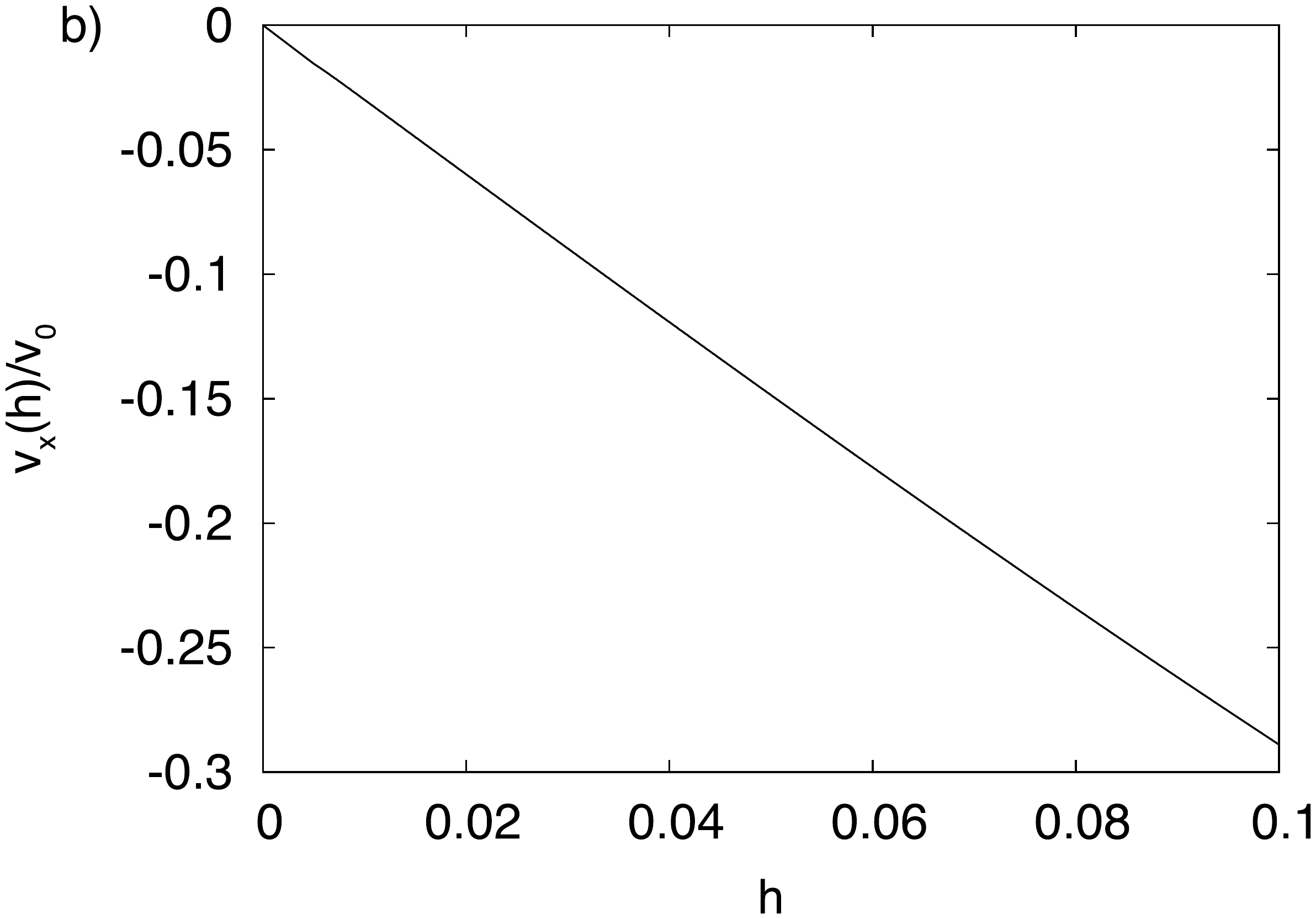}}
\caption{Velocity of swarm as a function of $y$.$h=0.02$, $x=0$ (a).
Velocity of swarm as a function of the relative thickness $h$. $x=0;~y=0$ (b).}
\label{Fig:1}
\end{figure}

The function $f(0,y,h-0,h)$ at $h=0.02$ is shown in Fig.~\ref{Fig:1}(a). For a comparison with the experimental observations, we calculate $v_{x}(h)/v_{0}=f(0,0,h-0,h)$ as a function of the dimensionless layer thickness, which is shown in Fig.~\ref{Fig:1}(b).

This model is tested against the previously described experiment. To receive the necessary constants, we describe the experiment in detail, while a detailed description of the microscopy platform can be found elsewhere \cite{12}. The cultivation medium reported by Heyen and Sch\"uler \cite{13} was used to grow the MSR-1 strain. A bacteria suspension with an optical density of 0.2 (NanoPhotometer\textsuperscript{TM} Pearl at 565 nm) was used to form MSR-1 swarms. The medium with the bacteria was degassed using nitrogen for 10 minutes and then inserted into a
rectangular micro-capillary (\# 3520-050, Vitrocubes, $T=200~\mu m$) by capillary forces. The loaded microcapillary was sealed at one end and left open at the other. The resulting oxygen gradient introduces a position with preferred oxygen concentration, where motile bacteria are concentrated \cite{14}. This process was accelerated by introducing a magnetic field antiparallel to the oxygen gradient to guide the swimming direction of the bacteria, which corresponds to an ideal case of magneto-aerotaxis in the environment \cite{7}. After 30 minutes the concentration of bacteria at this position was high enough to start the swarm experiments.

A rotating magnetic field parallel to the glass surface with a frequency of 1 Hz and a field strength of $30$~Oe was applied together with a constant magnetic field parallel to the normal of the glass surface with a field strength of $-30$~Oe. The combination of both fields together started the swarming behavior of the concentrated bacteria. 

A simple model of magnetotactic bacteria swarming under the action of rotating field is described in Supplemental Material \cite{15}. Since amphitrichous magnetotactic bacteria may act as pullers \cite{16}, we considered two bacteria rotating synchronously with the field \cite{17} and interacting due to flow fields produced by stresslets \cite{18}. This model as shown in figure Fig.~S1 in \cite{15} for hydrodynamic interaction parameter determined according to  the experimental data for flagellar motor of \textit{E-coli} bacteria \cite{19} 
predicts hydrodynamically bound states for interacting pullers. We may remark that similar mechanism of formation of hydrodynamically bound states  is described in \cite{20} for Volvox. Swarming of MTB in flow under the action of constant field along capillary axis is described in \cite{21}.

\begin{figure}
\centering{
\includegraphics[width=0.95\textwidth]{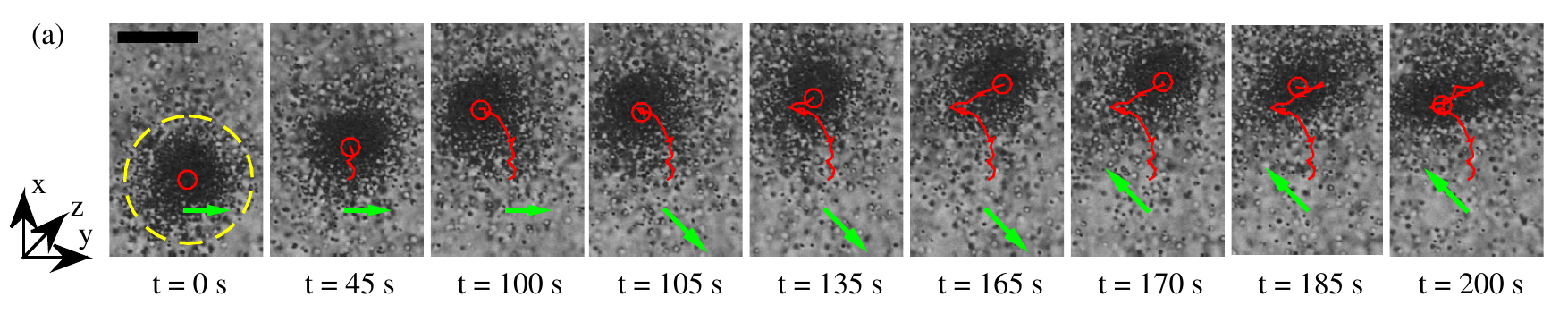} \\
\includegraphics[width=0.95\textwidth]{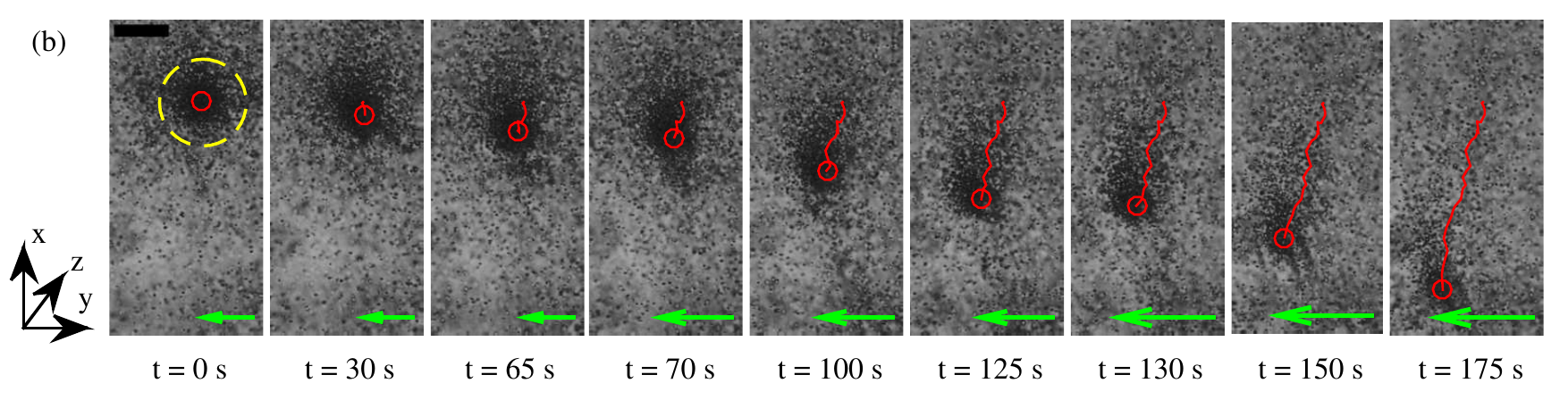}}
\caption{Motion of swarms due to an inclined external field. Red circles are swarm centers, red lines are centers trajectories. Green arrows visualize the magnitude of the in-plane magnetic field component. Yellow dashed line depicts swarm. $H_z=-30$~Oe at all times.
(a) Change of the in-plane component of the field alternates the direction of swarm motion.
(b) Increase of the magnitude of the in-plane field component speeds up the motion of swarm. Scale bars are $50~\mu$m.}
\label{Fig:3}
\end{figure}

After formation of the swarms, the rotating field was switched off, while the normal field remained $H_z=-30$~Oe. The swarms remained stable, as is visible in the first image of Fig.~\ref{Fig:3}~(a) and supplementary video S2 in \cite{15}. Swarms investigated here typically had a radius of $R\approx 40$~$\mu$m, as indicated with the yellow dashed circle. To investigate the potential of movement of the swarms on the glass surface, the constant magnetic field was inclined with an angle $\alpha$ with respect to the normal of the surface. 

Changing the direction of the in-plane magnetic field component $H_x$ or $H_y$, changes the direction of swarm motion, as shown in Fig.~\ref{Fig:3}~(a). At $t=0$~s the magnetic field $H_y=1.0$~Oe is turned on. At $t=105$~s additional $H_x=-1.0$~Oe, but at $t=170$~s, both in-plane components are inverted to $H_x=1.0$~Oe and $H_y=-1.0$~Oe.
This field inversion at $t=170$~s results in the motion of swarm in the opposite direction. If the magnitude of the in-plane magnetic field is increased, a faster motion of the swarm is observed, as visible in Fig.~\ref{Fig:3} (b) and supplementary video S3 in \cite{15}. In these images, at $t=0$~s, magnetic field $H_y=-2.0$~Oe is turned on, at $t=70$~s it is increased to $H_y=-3.0$~Oe, and at $t=130$~s it is increased up to $H_y=-4.0$~Oe. These observations are in agreement with relation (\ref{Eq:10}).

\begin{figure}
\centering{
\includegraphics[height=5cm]{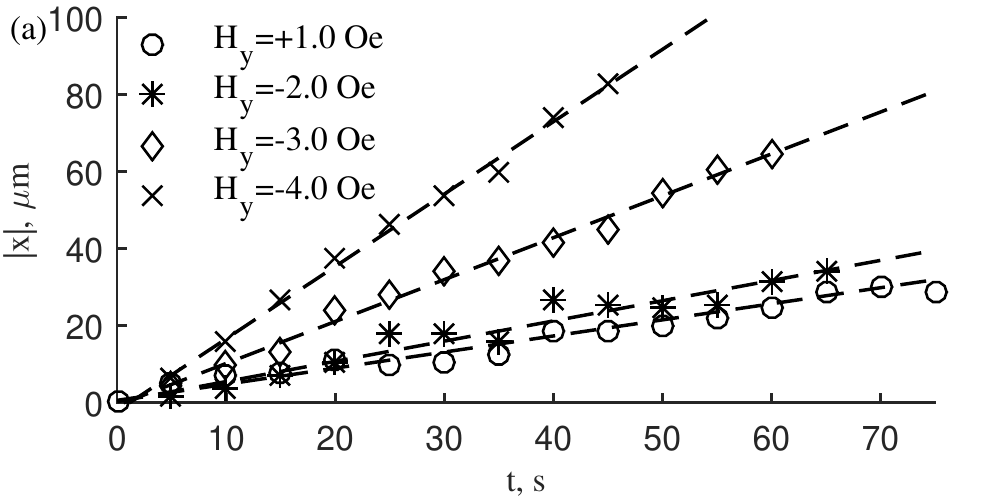}
\includegraphics[height=5cm]{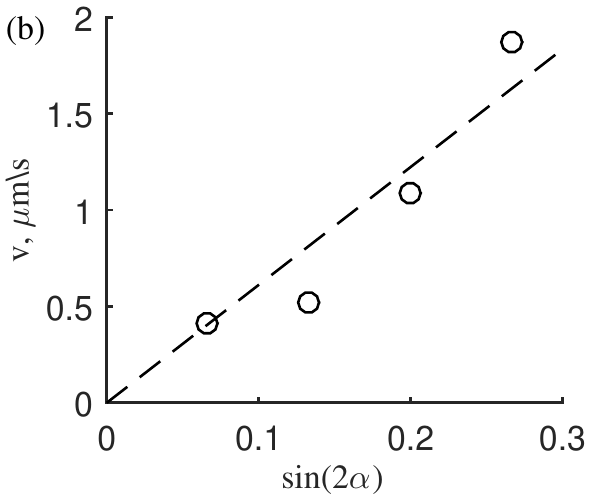}}
\caption{Velocities of the swarms. (a) Translation of the swarm as a function of time. Dashed lines mark best fits. (b) Swarm velocity as a function of the sine of $2\alpha$. Dashed line is the best fit, with a slope $v_{*}=6.1~\mu m\cdot s^{-1}$.}
\label{Fig:4}
\end{figure}

To quantify swarm motion, we tracked the motion of swarm centers. Due to a dynamic field of view, tracking is done manually. We analyze the translation of the swarms in the direction perpendicular to the in-plane magnetic field.
Results for several in-plane magnetic field magnitudes are shown in Fig.~\ref{Fig:4}. The translation motion is close to linear: at $H_y=1.0$~Oe $|v_{x}|=0.42~\mu m\cdot s^{-1}$, $H_y=-2.0$~Oe $|v_{x}|=0.52 ~\mu m\cdot s^{-1}$, $H_y=-3.0~ Oe$ gives $1.09~\mu m\cdot s^{-1}$ and $H_y=-4.0~Oe$ provides $1.88~\mu m\cdot s^{-1}$. The speeds are determined by a linear fit (dashed lines).

Swarm velocities increased with the in-plane magnetic field. We estimated the magnetic bacteria concentration by approximating with a single layer bacteria disk (see yellow dashed circle in the first image of Fig.~\ref{Fig:3}~(a)) and thresholding the circular swarm area (\textit{ImageJ}). A reasonable concentration of $n_{\Sigma}\approx 0.5~\mu$m$^{-2}$ was determined. It should be remarked that relation (\ref{Eq:10}) describes the direction of swarm motion at $\tau>0$. For bacteria stuck at the lower wall at $z=0$ (Fig.~\ref{Fig:3}(b)) $l_{y}<0;~l_{z}<0$ and since $f<0$  (Fig.~\ref{Fig:1} (b)), the swarm moves in the negative $x$-axis direction (Fig.~\ref{Fig:3}(b)).

Swarm velocity is in agreement with relation (\ref{Eq:10}) and is well described by $v_{x}=v_{*}\sin{(2\alpha)}$, where $v_{*}=6.1~\mu m/s$. The obtained  value may be compared with the value from relation (\ref{Eq:10}). Since the velocity profile given by relation (\ref{Eq:10}) is linear and the dimensionless thickness of the layer may be estimated as $h=0.02$
then $k=n_{\Sigma}\tau f(0,0,h-0,h)/16\eta$, where the volume concentration of bacteria is estimated as $n=n_{\Sigma}/h$ and  $b$ is put equal to the thickness of the layer. Taking for the torque developed by the rotary motor the value $5~pN\cdot\mu m$  \cite{22} and $f(0,0,0.02-0,0.02)=-0.06$ we have $k=9.4~\mu m/s$, which is close to the experimental value $v_*=6.1~\mu m/s$ . Some discrepancy may be caused by the overestimation of $b$ and the fact that boundary of the swarm is not as sharp as supposed in the calculation. Let us remark that the sign of $\tau$ corresponds to the sign assumed in the model of the formation of hydrodynamically bound states described in the Supplemental Material \cite{15}.

To sum up we showed that collective motion of swarms is initiated in ensembles of magnetotactic bacteria with orientation order. The formation of swarms under the action of a rotating field is described as arising from hydrodynamically bound states due to hydrodynamic interaction produced by stresslets.  The motion of these swarms  is perpendicular to the plane defined by the normal of the surface and the oblique field and may be described by a continuum model considering the couple stress due to the torque dipoles of bacteria. Quantitative agreement of the predictions of the model with experimentally measured swarm velocities is obtained using a known value of the torque produced by the rotary motors of bacteria.
\section{Acknowledgements}
This study has been carried out during a project "Stochastic dynamics of magnetotactic bacteria at random switching of rotary motors" of the Baltic-German University Liaison Office, which is supported by the German Academic Exchange Service (DAAD) with funds from the Foreign Office of the Federal Republic Germany. Authors are grateful  to Rudolfs Livanovics for fruitful discussions.

\end{document}